\documentclass[aps,twocolumn,epsf,floats,prl,reprint,nofootinbib]{revtex4-1}
\usepackage{graphics,graphicx,epsfig}

\begin{document}
\title{Life at the edge: complexity and criticality in biological function} %
\author{Dante R. Chialvo}
\affiliation{Center for Complex Systems \& Brain Sciences (CEMSC$^3$), Escuela de Ciencia y Tecnolog\'ia. Universidad Nacional de San Mart\'{i}n (UNSAM),
25 de Mayo 1169, San Mart\'{i}n, (1650), Buenos Aires, Argentina}
\affiliation{Consejo Nacional de Investigaciones Cient\'{i}ficas y Tecnol\'{o}gicas (CONICET), Godoy Cruz 2290, Buenos Aires, Argentina}

\begin{abstract}
Why life is complex and --most importantly-- what is the origin of the over abundance of complexity in nature? This is a fundamental scientific question which, paraphrasing the late Per Bak,  ``is screaming to be answered but seldom is even being asked''. In these lectures we review recent attempts across several scales to understand the origins of complex biological problems from the perspective of critical phenomena. To illustrate the approach three cases are discussed, namely the large scale brain dynamics, the characterisation of spontaneous fluctuations of proteins and the physiological complexity of the cell mitochondria network. 
\end{abstract}
\maketitle
\section{Introduction}
  
In the last decade we have witnessed an escalating interest in complex biological phenomena at all levels including macroevolution, neuroscience at different scales and molecular biology.  Potential progress is of paramount importance, thus  we shall examine a bit how we are currently proceeding to carve these new areas, starting with asking whether  biological phenomena are more or less complex than other fundamental problems in physics. The answer is not clear at first, however striking differences exist in the approaches as well as in the sociology of both fields.

The history of physics records many important efforts in search for universality, large classes of phenomena must be explained in terms of a few fundamental laws.  In contrast biology more often seems to emphasise unique and singular aspects; because not all organisms are alike, there is plenty of diversity of species, families, etc. such that taxonomy ends up prevailing over integration of knowledge. This apparent uniqueness of each biological phenomena in some cases leeds to overspecialisation, which may from time to time encourage the creation of a sub-discipline for each new group of complex biological phenomena. Of course, this tendency prevents fruitful dialogue between biology and the rest of the sciences, thus leading to an {\it exponential  increase of our knowledge about almost nothing}, or in other words to the fragmentation of the biological scientific inquire into many disconnected ``cottage industries". 
 
Often it is also argued that biology could not be well studied by physics, because ``the laws of physics are simple but nature is complex''.  This is motivated by the assumption that anything that ``looks'' complex originates from laws that must also be complex. Thus, the idea that has been perpetuated is that the complexity of nature is almost inaccessible, arguing that the diversity and ever changing fluctuations shown by natural objects prevents their study through mathematical tools. In contrast, others called attention to the fact that \cite{glass}

\begin{quote}  ``\it ... if the complex dynamic phenomena that occur in the human body were to arise in some inanimate physical system --let us say  in a laser, or liquid helium or a semiconductor--  they would be subjected to the most sophisticated experimental and theoretical study.''
\end{quote}

These lecture notes\footnote{Lectures notes from an introductory tutorial given at  the LVIII Cracow School of Theoretical Physics ``Neuroscience: Machine learning meets fundamental theory'', June 15--23, Zakopane (Poland).}  adhere to the spirit of the above quote and aim to illustrate some successful attempts to study complex collective phenomena \cite{chialvo} with approaches borrowed from statistical physics. Rather than going into the details of each of the studies reviewed, the emphasis here will be to dwell in the logic behind adopting this approach to study biological function. Another  cautionary note is that we are here preaching for the {\it non-cognoscenti}, and being the topic at the fringe of disciplines, surely physicists and biologists alike will encounter boring passages on their most familiar topics.  

The next sections will progressively introduce the problem of complexity (in Section 2) and how its origin can be related to critical phenomena. The examples were chosen with the intention to persuade the reader that the same simple laws apply exactly to very different complex phenomena, a notion known in physics as universality. 
After defining the issues, our own advances in the use of this approach to study complexity in life will be discussed by presenting three problems, starting with the  description of our long-going work on brain dynamics \cite{chialvo} in Section 3.  Then the issue of protein dynamics will be discussed in Section 4 by reviewing our recent work \cite{tang} on the finite size scaling analysis of structural protein data from a large database. After that, in Section 5, we review an empirical and theoretical analysis \cite{nahuel1} able to uncover a critical fusion-fision balance in the mitochondrial network of a cell. The paper concludes with a summary of the main message.

\begin{figure} [!hb]
\centerline{
\includegraphics[width=.9\linewidth]{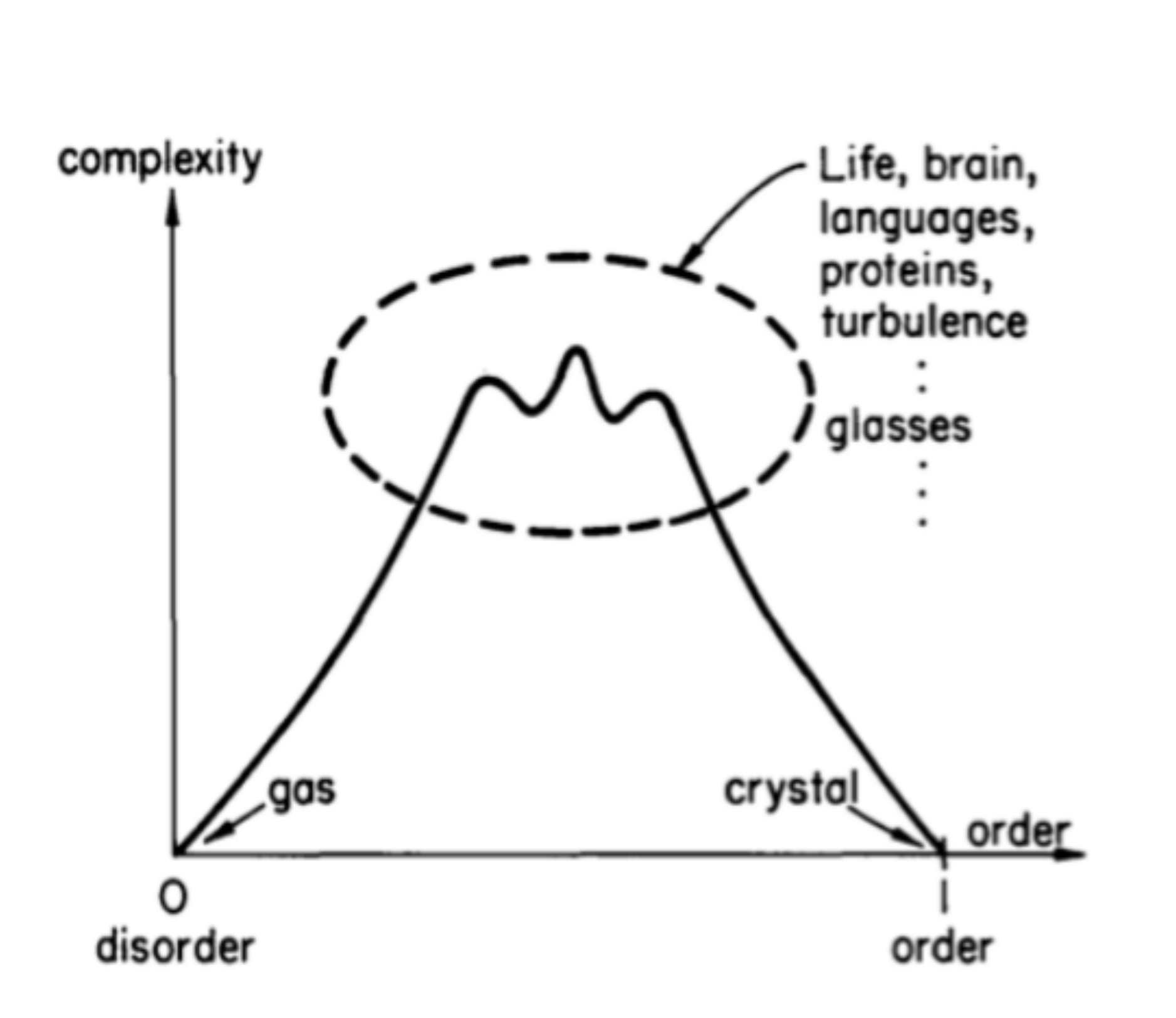}}
\caption{\label{intro1}Neither the excessive disorder of a gas nor the extreme order of the molecules of a solid are perceived as complex. Generally, complexity is perceived as having  intermediate levels of order and disorder, as illustrated in this cartoon four decades ago by Hans Frauenfelder. It is in this intermediate region --exhibiting a mixture of order and anarchy-- where the most complex phenomena inhabit, including life, language, proteins, turbulence, glassy states, etc. (From Frauenfelder \cite{hans})} 
\end{figure}

\section{Complexity}

\subsection{Recipe for natural complexity: a bit of order and another bit of disorder}
Given the fact that complexity is ubiquitous in nature is then obvious to wonder about how it is built. It has long been suspected that the answer to this question lies at the border between order and chaos. As denoted in Frauenfelder\cite{hans} cartoon of Fig.\ref{intro1} a bewildering variety of apparently disconnected phenomena --all of them dubbed complex--  exhibited an intermediate level of order and disorder; including life itself, brain, languages, proteins, turbulence in fluids, the slow dynamics of glasses, to name only a few.

Clearly, something  repetitive  (as in the extreme order) does not seem difficult to explore, as would be the case of a crystalline structure. In the same way, what does change erratically  in anarchy, as is the case of the trajectories of the molecules of a gas, does not look complex. On the other hand, something that occasionally stops being monotone (whether in space or in time) surprises us and becomes something intriguing and complex. That fair and balanced mixture of order and disorder, or surprise and boredom is commonly the letter of presentation of complexity. 

Everyday examples abound, let's take the case of music where there is a balance between surprise and repetition, avoiding excessive monotony or frequent surprise. Another example, involving spatial aspects, could be fingerprints, all similar and different at the same time. We could ask ourselves if the complexity of the mixture we observe is related to the complexity of the mechanism that generates it. In other words: must we assume that to manufacture the precise ``mixture''  that prevails in something complex,  requires of new and more complex laws than those necessary to generate the extreme order or the disorder?  We will show that the same simple laws can explain the simple and the complex. 

\subsection{Phases and universality} Perhaps being a daily experience, we fail to notice that matter in nature  comes to us in a few ``phases'' or states, for example water, mostly in three: It is important to note that in spite of the great qualitative differences between the three states, exactly the same physical laws govern the behavior of their constituent molecules.  A relatively small change,  in temperature or pressure, can originate very different collective behaviors of the same molecules. In other words, monumental collective changes, which are reflected as different phases, do not require different molecules, not physical laws, neither any fundamental change in the laws ruling the molecules ``interactions''. Lets inspect the case of water: vapor is a gas at the macroscopic level and if we observe it with a powerful microscope we could count billions of water molecules moving crazily in any direction, at more speed the greater the temperature of the vapor.  If we slowly cool this gas, we will see that the same molecules move slower, and that small groups begin to form. This occurs because, as the temperature decreases, the mutual attractions between the molecules begin to overcome the tendency to disorder that the thermal agitation gives to them and the molecules tend to come together.  Soon, the small initial clusters continue to capture other molecules, forming drops of water, when the temperature is below 100 degrees centigrade. If the temperature continues to drop the attractive forces between molecules begin to play an increasingly important role in opposing the thermal agitation and at 0 degrees, they will be able to produce regular microscopic structures, thus causing the solidification of water into ice. These two changes (condensation or solidification and vice versa) are called in physics phase changes (or phase transitions). 

Not more than a century ago, it was thought that these changes -now described as phase transitions- mean different matters; a replacement of one thing for another, steam for water and this later for ice, because matter was considered to be continuous. This vision continued until, at the dawn of the twentieth century, it was confirmed that matter were sets of atoms and thus it became clear that despite large qualitative differences in the appearance of the phases they involve the same molecules changing only their conformation. It is interesting to note that coincidentally, Ramon y Cajal also broke the existing idea that the brain was a syncytium, histologically identifying the synapse and then demonstrating the discrete nature of the nervous system. 

Phase transitions occur in all the matter that surrounds us, and its study has been systematised recently in a great variety of collective phenomena that occur whenever a large number of non-linear elements interact. 
It is known, for example, that the correlations between the parts that make up a system obey statistically identical rules, regardless of whether the constituent elements are neurons, ants, grains of sand or water molecules. In all cases, the same theory explains how the system is ordered or disordered, what types of collective behavior can be expected, how stable or unstable they will be, how it can be disturbed etc. The fact that all these disparate phenomena obey the same laws is what is known in physics as universality. 
\begin{figure} [ht] 
\centerline{
\includegraphics[width=.95\linewidth]{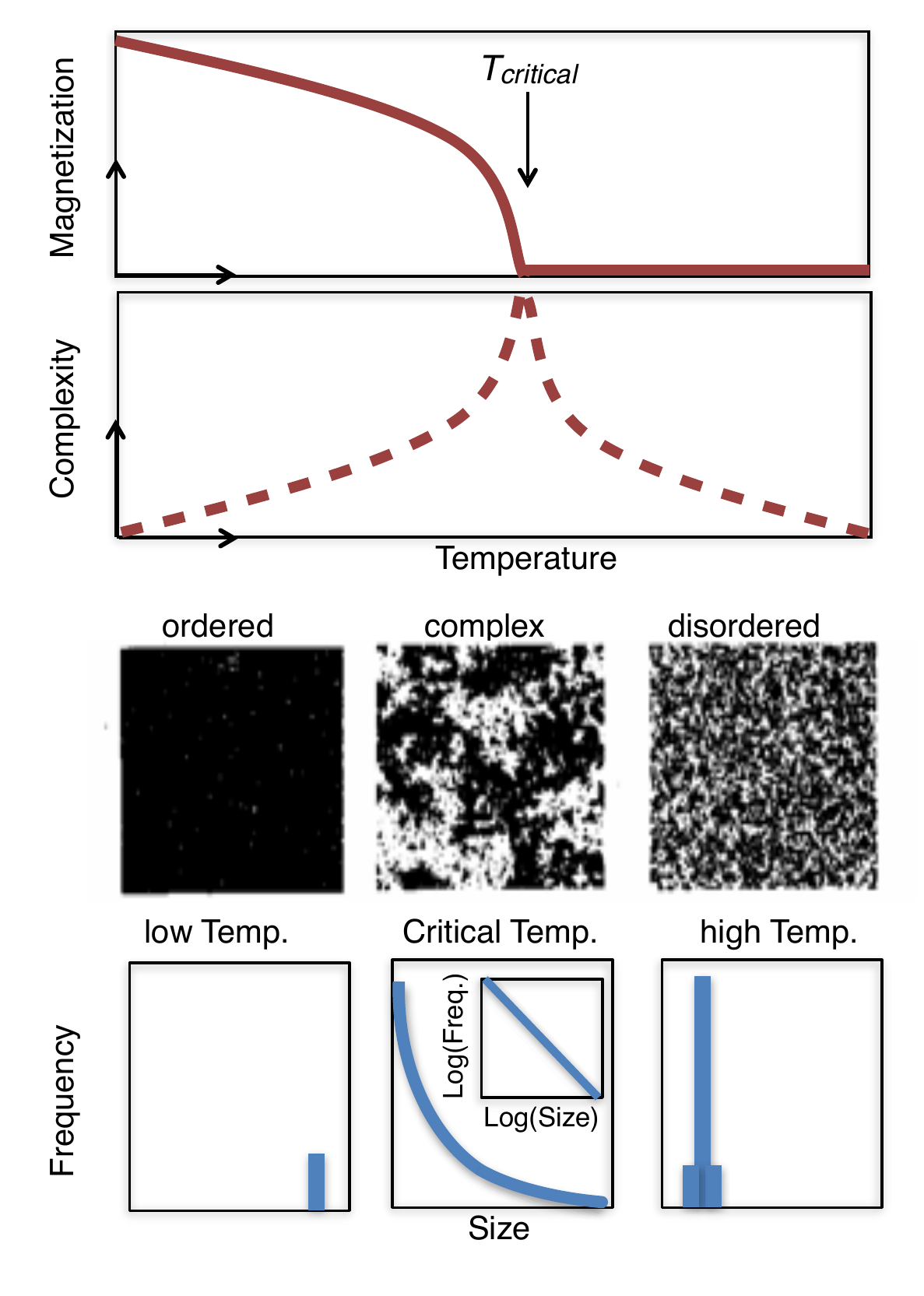} }
\caption{\label{intro2} Example of a phase transition, one of the most frequent mechanism that generate complexity in nature. The two upper panels illustrate the change in the magnetization and in the expected complexity of a ferromagnetic material as a function of temperature. Below, examples for the  three phases of the system: ordered (low temperature), disordered (high temperature) and close to the critical temperature (complex). The lower graphs illustrate the distribution of the size of the ``islands'' of equal orientation (i.e., those with the same color), which is very homogeneous for extreme temperatures, but it is scale-free close to the critical temperature. Complex systems by definition show this type of scale free distributions, which when plotted in double logarithm axis (as in the insert diagram) result in a straight line.}
\end{figure} 

To accept that the same laws govern and explain apparently very disparate phenomena, is a process of generalization not without difficulties. It is enough to imagine Galileo Galilei trying to persuade the theologians that the celestial bodies were governed by the same laws as a vulgar stone or a bird feather. It was obvious that they would protest,  ``How to pretend that those majestic celestial bodies circulating the heavenly spaces where the gods reign will follow the same rules than these mundane objects? Today, rationale of using the exact same laws to describe the oscillations of a swing and the evolution of planets in its orbits is easily admitted, but still only a minority is inclined to accept that the laws of physics must be fundamental to understand the world of neuroscience. 
This explains the reluctance to admit that the interactions between a multitude of neurons can trigger collective phenomenologies that are qualitatively equivalent to those we observe, for example, as a product of the interaction between atoms of a metal. 
 
\subsection{Complexity arises in between order and disorder} To describe the universal scenario of the complexity that emerges at a phase transition we will consider the prototypical case of magnetization, an example of collective phenomena. The cartoon on Fig.\ref{intro2} shows the behavior of a piece of iron subjected to an external magnetic field as the temperature increases. Without going into much detail, the atoms tend to align their magnetic moments with those of their immediate neighbors. In turn, this tendency to order competes with the agitation that temperature produces. If the temperature is low, the final state of the system will be ordered with all spins oriented in the same direction.  

The degree of order-disorder of the system can be followed by choosing the appropriate variable, in this case is the magnetization. This is maximum when order prevails (that is, where the image acquires the configuration that is familiar to us: with a north and a south pole) and vanishes when disorder prevails, that is when the neighbouring magnetic moments are randomly oriented.

The examples in the three intermediate panels in Fig.\ref{intro2} show an instantaneous image of the state of the system, where black/white represents north/south spin orientation. It can be seen that at very low temperatures almost all spins coincide, meaning that order prevails; while at very high temperatures disorder prevails resulting in alternating small neighbouring regions with opposite alignments. Although the spatial patterns we see are different, they are homogeneous through out the system. The complexity of these patterns can be evaluated in many ways, for instance one of them involves algorithmic complexity, estimated by computing the length of the algorithm needed to describe that state. If the pattern to be evaluated is repetitive and homogenous - as in extreme temperature - then the complexity will be vanishingly small. On the other hand, complexity is expected to be high at temperatures close to the critical point, since the spatial patterns  correspond to non-homogenous and complex mixtures of disorder and order. 

The patterns at critical temperature show a great deal of heterogeneity: there are black ``islands'' (indicating a coincidence in the spins orientation) of all sizes, which in turn contain white lagoons, which are also of all sizes. This is contrary to what is observed at the extremes, close to the critical point there is no preferred size for island or lagoons, in fact the observed patterns are ``scale free". Scale-invariance also implies the largest complexity, because it means the largest number of configurations. This absence of scale results in a continuous function  obeying a power law, $ P (S) \sim 1 / S ^ \beta $, where $ \beta $ is the exponent that characterize the distribution of sizes $ S $.  This type of function is distinctive of the behavior of complex systems, and it is easily recognized when doing logarithms in both axes a straight line is perceived, as illustrated in the bottom central panel of the Fig.\ref{intro2}. 

Complexity can also be observed in the time domain analysing the temporal fluctuations of the order parameter. The magnetization as a function of time at both extremes of temperature exhibits very small fluctuations, while close to the critical point shows  episodes of apparent calm  that are interrupted from time to time by large variations. The variability of the magnetization over time is also ``scale-free'', a consequence of the fact that in complex systems the spatial and temporal dynamics are not independent, they are two sides of the same coin. 

The universality discussed here suggests that the way in which complexity emerges in the example of the magnetization can be seen generically in phase transitions  at systems very different from one another. Indeed many examples can be found in the recent literature  such as birds flocks, large groups of neurons,  stockbrokers interacting, etc.  We will discuss three examples including important aspect of cerebral dynamics, as well as proteins and mitochondrial dynamics, all  governed by common universal principles. 

\subsection{Complexity is always critical} The preceding paragraphs summarize one of the lessons of statistical physics: complexity and criticality are almost synonymous: what makes a system complex are exactly the same  properties exhibited by a system when it approaches the critical point of an order-disorder phase transition. (Let the reader ignore for the moment how a given system manages to reach criticality.) The main point is that close to the critical point the spatial patterns exhibit a mixture of order and disorder: not all the microscopic elements of the system do the same, nor do each one behave randomly. In this way, the ``repertoire" of patterns that the system is able to exhibit increases.

Turning to the biological consequences of criticality, note that the combination of collective tendencies of order and disorder is fundamental for the adaptability of any collective: it needs a certain regularity to function, but also it must be flexible and variable in order to adapt to changes in its environment. For instance  let's think about the brain case: If all the neurons behaved suddenly in the same way, we would be witnessing an epileptic attack.  In the other extreme, if each neuron behaved randomly, they will be uncorrelated and there would be no exchange of information making impossible any concerted output. In both cases, extreme order or extreme disorder, it is inconceivable that the brain could work. 

\begin{figure} [ht] 
\centerline{
\includegraphics [width=.95\linewidth] {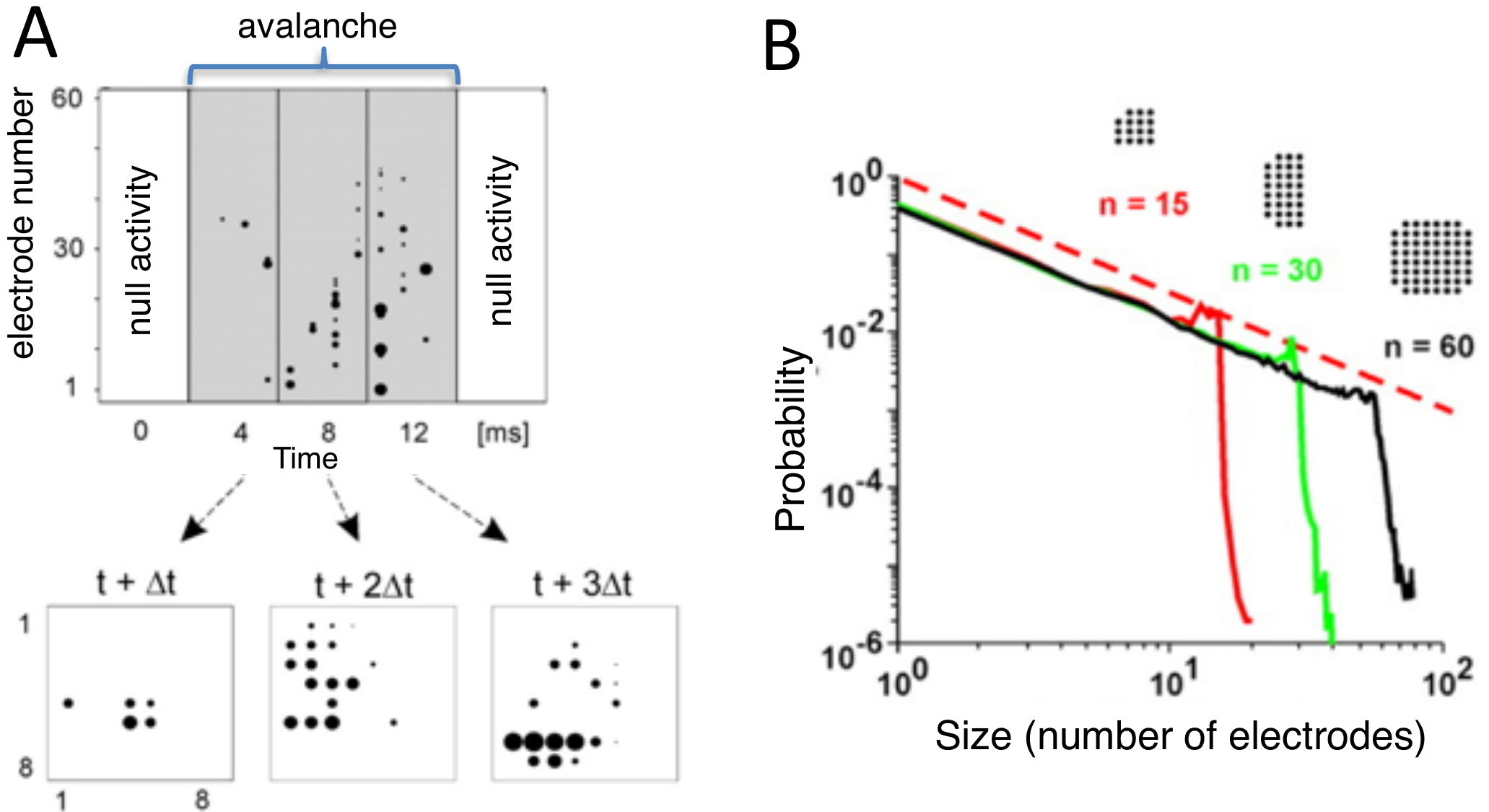} }
\caption{\label{brain1}Avalanches of neural activity are critical (and complex): Panel A: Avalanches are clusters of neuronal discharges separated by periods of silence (in the upper panel illustrated with points for each one of the 64 electrodes arranged in an 8x8 grid). The avalanches' duration and size are very variable. In this example the avalanche persists for 12 ms traveling trough 38 electrodes (number that defines its size), as shown by the sequence in the three lower panels which shows the location of the active electrodes in that step of time. Panel B: The distribution of the avalanche size follows a power law (dotted line). The relative probability of an avalanche of a given size is plotted. The size of the biggest avalanches is only limited by the size of the system, as evidenced by the three examples recorded in systems of 15, 30 or 60 electrodes. Redrawn from Beggs and Plenz \cite{beggs}.} 
\end{figure}

\section{The brain: could it be that its laws are simple?} 
The complexity of the brain fascinates everyone and sometimes it is argued that in such complexity lies our mere inability to understand its functioning. Instead, a naive approach seemed more attractive to us. Beginning with the initial works in the 90's in collaboration with the late Danish physicist Per Bak, we proposed to look at the brain in terms of phase changes and critical dynamics \cite{bak} as if the problem were any other physical phenomena.  Our small steps received a significant boost, in 2003, by the Beggs and Plenz \cite{beggs} experiments which  provided the first clear evidence of critical dynamics in neural data. They described, in cultures of neurons, a phenomenon that they dubbed ``neuronal avalanches'',  a spatial pattern of electrical activation of the cerebral cortex in which cascades of activity with peculiar statistics are propagated by the whole system. In this work the experimental setup allowed to follow the propagation of the neural events through a grid of electrodes, which recorded any activation in their vicinity (Fig.\ref{brain1}A). When the sizes of these avalanches were studied, they found that they do not exhibit any preferred value, they were, as in the example of magnetism, scale-free. That is, when these avalanches are plotted on double logarithm axes (see Fig.\ref{brain1}B), a function that follows a straight line is outlined. This finding, and its subsequent replications in various conditions, sparked great interest and prompted research on the subject, across animal, modelling and human scales.
 
To be brief we will only refer to human experiments with neuroimaging techniques, in which we have participated more actively.  These recording techniques use functional magnetic resonance (fMRI), which measures brain activity indirectly by detecting changes in blood oxygenation, which is associated with the metabolic consumption produced by the neuronal activity. The data obtained every second with this technique represents an image of the whole brain parcelated in thousands of cubes of a few cubic millimetres (the so-called voxels).  

Using this type of data we have explored how the resting brain continuously approaches and moves away from the critical point. It is obvious to ask what we could learn by observing a brain that is ``doing nothing'', i.e., resting? The answer is in the spirit of the fluctuation-dissipation theorem: the magnitude of the variance of the spontaneous fluctuations in a complex system is nearly proportional to the response to an external perturbation. 
We have investigated the properties of the brain spontaneous fluctuations  by extracting  the moments in which the signal of the functional magnetic resonance exceeds a given threshold \cite{PP1,PP2,PP3}. This transformation generates a ``point process'' which represent the relatively large events in time and space. In spite of being an extreme simplification of the original time series, it has been shown that is not accompanied by loss of information. Furthermore the point process allows to follow then instantaneous brain activity continuously with great fidelity. With this technique it is possible to describe the brain dynamics following the evolution of these points in time and space, as if they were stars in the sky. In analogy, we can look at the correlation properties of these points, as if there were ``constellations'', where, how many, what sizes, how they move, etc. 
\begin{figure} [ht] 
\centerline{
\includegraphics [width=.95\linewidth] {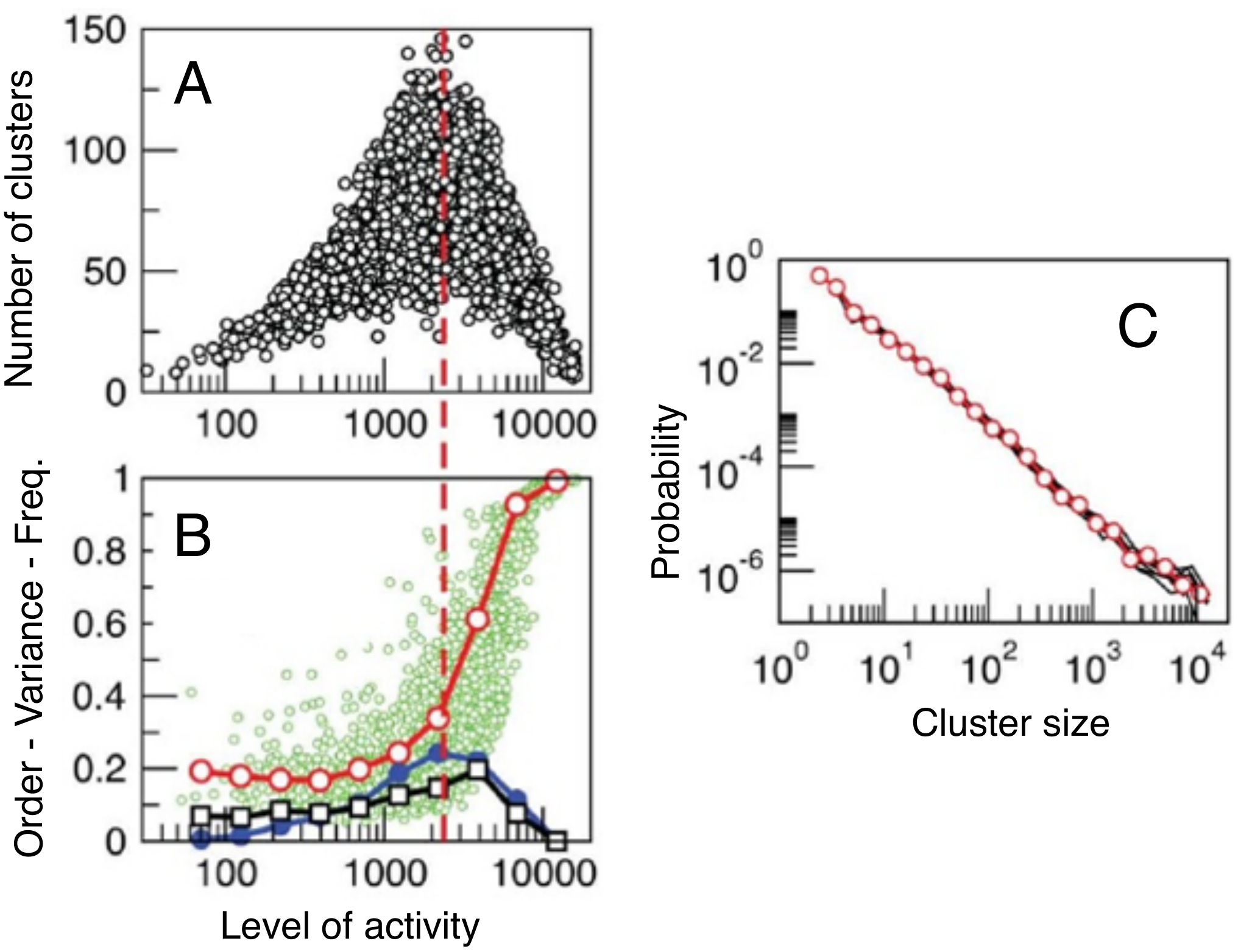} }
\caption{\label{brain2}Brain activity recorded using functional magnetic resonance imaging exhibits fluctuations around a phase transition.  Panel A: The number of clusters of brain activity (vertical axis) as a function of  the level of activity shows a maximum (denoted by the vertical dotted line). Panel B: Order parameter (green symbols and its average indicated by empty circles connected with continuous red line) and its variance (squares and continuous black line) as a function of the level of activity. The filled circles (whose maximum corresponds to that of the data in panel A) correspond to the frequency in which a given level of activity is observed. Panel C: The probability density distribution of the brain activity cluster sizes is scale-free. Re-drawn from \cite{PP3}. } 
\end{figure} 
 
This type of analysis revealed for the first time a series of very interesting properties of the large scale brain dynamics \cite{PP3}, some of them summarized in Fig.\ref{brain2}. At each time step the total number of points (i.e., activated brain sites) represents the degree of brain activity at that time. Looking at the spatial distribution of the points the level of clustering can be estimated. Upon analysis it was immediately apparent that the number of clusters and its size fluctuates in time, following a familiar pattern. As shown in panel A of Fig.\ref{brain2}, the number of clusters obeys a non-monotonic relation with the number of activated sites. In addition the variability of the number of clusters is larger near the peak of such relation.

This simple analysis suggests that the brain activity is always close to a phase transition. To make these considerations quantitative, and to compare them with other systems, an order parameter was defined and  calculated from the data.  In this case the order parameter is equivalent to the magnetization commented for the example in Figure 2. Here it is defined (at each instant) as the size of the largest cluster of actives sites. In turn, the activity level, already defined, can be considered as a pseudo-control parameter (equivalent to the temperature in Figure 2). Now, by plotting the order parameter as a function of the control parameter, the sigmoid curve  of panel B  of Fig.\ref{brain2} appears, which suggests the existence of a phase change when the level of activity increases. Confirming this indication, the variance of the order parameter shows a maximum that locates the point of the possible phase transition. Remember that the greatest variability is often observed near criticality. Given that the level of activity fluctuates three orders of magnitude, it is appropriate to ask how often the brain is close to criticality. This is done by measuring the frequency with which the system is at each level of activity. We found that, in effect, the brain spends relatively more time (see filled circles in Fig.\ref{brain2}B) around a transition zone of intermediate level of activity. Finally, the graph of Fig.\ref{brain2}C shows that the statistics of the size of the clusters of activity follows a power law, which is characteristic of criticality, as shown for the example of magnetism in Fig.\ref{intro2}. 

The behavior described by the inverted-u-shaped curve in Fig.\ref{brain2}A is very common in physics, being another manifestation  of universality. In the studies of road traffic, the same functional form is observed for  the relation of the flow of vehicles passing through a control site as a function of the density of vehicles occupying a given section of the road.
As in Fig.\ref{brain2}A, it is typical to see that for relatively low densities, the flow of vehicles initially grows proportional to the density of cars.  That is, as more vehicles enter the road, the flow through a given point increases. But for a  relatively high density the flow reaches a maximum,  at which point the odious traffic jams occur.
Moreover the variability of the traffic is maximum close to the critical density, that is, the time it takes to travel the same route in different days become highly variable.

Thus the curves in Fig.\ref{brain2}A and B show that the brain spontaneously fluctuates between two extremes,  one with low activity, where there are only a few small clusters (such as a clear sky with a few small clouds) and the other with high activity (dominated by a huge cluster -like overcast sky-). 
 
\subsection{Consciousness}  
\begin{figure} [htbp] 
 \centerline{
 \includegraphics [width=.95\linewidth] {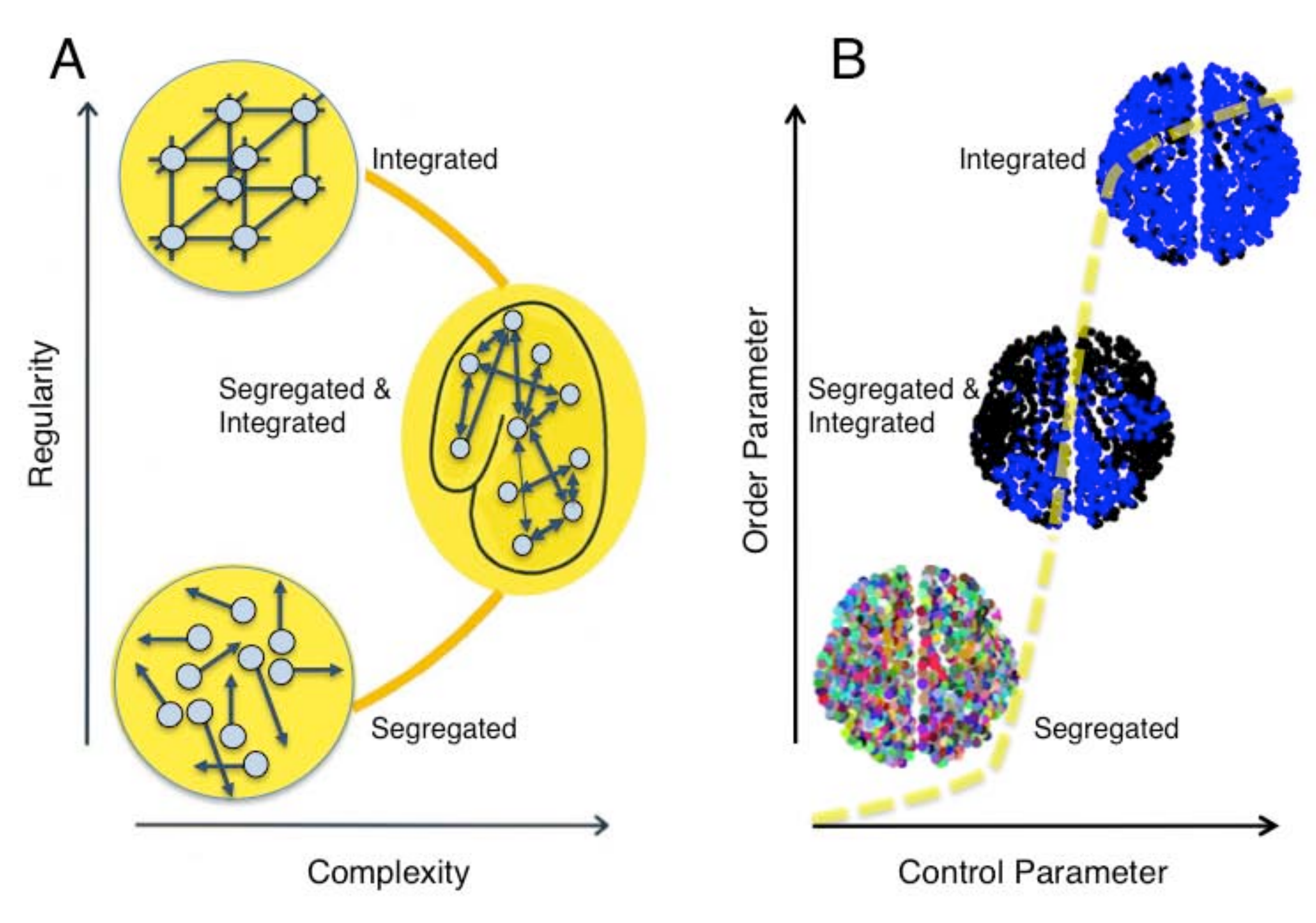} }
 \caption{\label{brain3}According to the predominant theories, the conscious state is characterized by the capacity of ``maximum integration--segregation'' of information. Panel A shows the structural point of view used by Tononi in which different degrees of regularity in the interactions (or connections) of a neural network can confer more or less capacity for segregation or integration. Panel B shows our dynamic alternative by which, in principle, any of the three regimes - with more or less complexity - can be generated dynamically by a system (without any change in connections) able to undergo a continuous phase transition. The various colors in the three graphs identify the different clusters of activity, so that in the most ordered one (top right) the entire brain is active, while in the more disordered (bottom left) each region acts independently. Coherent clusters of all sizes can be seen only at criticality (middle graph), a condition at which  is obtained the best possible integration--segregation balance is obtained. Adapted from \cite{mentecritica}.} 
 \end{figure}

Defined as ``...that thing which disappears in deep sleep (when on awakening we can not report where we were) and which reappears as we woke up", consciousness is hard to formalize. Giulio Tononi is perhaps whom has worked harder into quantifying  the subjective aspects of human consciousness, through ingenious experiments and theoretical arguments \cite{tononi,tononi2}. While admitting that it exists only in ``first person'' his central theoretical argument establishes that consciousness is a state where the {\it ability to simultaneously integrate and segregate} information is maximum. The simultaneity of these opposing properties could appears as a contradiction, however this coexistence, according with the theory,  is necessary to explain the most fundamental properties of conscious experience.  In its original formulation  Tononi  imagines the interactions in the brain as in the three phases or states of mater, one very segregated, one very integrated and the intermediate, which contains a mixture of segregation and integration. The last one  corresponds to the conscious state (see Fig.\ref{brain3}A). This formalism does not need to open any judgment as to how such coexistence is achieved; in other words, it does not propose a specific neuronal mechanism which may endow  the brain with such properties. 

A decade ago we already noticed \cite{edge} that there is a striking similarity between the extremes of segregation or integration and the gas and solid states of matter. It was apparent to us that  the critical state, intermediate to these extremes, meets the conditions required by Tononi for the conscious state in his ``Integrated Information Theory''.  Then, we have proposed that the solution to the problem of how different degrees of integration and/or segregation are achieved in the brain is not related to changes on the {\it interactions} but with changes in the {\it correlations}, such as what happen in a typical second order phase transition. According to this notion, the structure of the brain connectivity, (i.e., the interactions) can be --at relatively fast time scales-- immutable, nevertheless the conscious state can be easily manipulated by adjusting a single parameter. The ability of the {\it same structure of interactions to exhibit completely different correlations} resembles similar changes in phases demonstrated by the examples of water and ferromagnetism. Note that Tegmark \cite{Tegmark} reached independently similar conclusions, starting from earlier complementary considerations \cite{tegmark1}.

According with our perspective, the conscious state correspond to the critical state, as illustrated by the cartoon in Panel B of Fig.\ref{brain3}.  It is important to note that we are not implying (at all) that any  system that is critical is also conscious, but that to be critical is a necessary condition for a system to exhibit the segregation-integration properties advocated by Tononi as crucial to endow a system with consciousness. The implication of this relation deserves further exploration.

\subsection{New directions} The results of this line of work helps to understand the spatiotemporal dynamics of the brain, even  when it is doing nothing in resting conditions. The universality that we comment, allows us to cross borders and include funny analogies like ``brain meteorology''  when we describe these studies. In that sense, an euphemism that we allow ourselves is to say that knowing brain weather patterns in healthy conditions can allow us to understand how pathological storms, droughts, etc., can occur and how to proceed to recover a healthy brain weather. 

Fundamental results usually give rise to more questions than answers. In this sense, in addition to the study of the healthy brain under the optics of statistical physics, we have recently investigated brain integrity in various physio-pathological conditions. For example, our most recent work was dedicated to study functional magnetic resonance records in human volunteers with different degrees of consciousness which were found  to be analogous to the qualitative changes observed in the phase transitions already discussed here \cite{propofol}.
 
From that perspective, the state of vigilance would correspond to the critical state, whereas the deep sleep or the loss of consciousness due to general anaesthesia are consistent with a sub-critical state. On the other hand, we have proposed that the alterations of the consciousness produced by hallucinogens would correspond to a supercritical state if we consider that the entropy of states is increased \cite{hongos,robin}.

A special mention must be made to the mathematical modelling of these results \cite{haimovici}. In that work a model was constructed by using  the empirical connectivity data between regions of the human brain (obtained from the so-called Human Connectome project \cite{connectome}) and the neuronal dynamics described by  a very simple non-linear dynamic rule. These results were able to show that the totality of the resting state cerebral dynamics observed experimentally can be replicated, just by tuning the model to a region close to the critical point. In contrast, when the exact same model was barely mistuned outside criticality failed to replicate the well known dynamics of the resting state networks, thus highlighting our idea that brain dynamics emerges only close to criticality \cite{fraiman}. The model was recently revisited by Rocha et al. \cite{rocha} to investigate how re-normalization of the local interactions may affects the macroscopic activity during rest and the formation of functional networks.

Overall these results open up the fascinating possibility of constructing and exploring ``virtual'' computational brains, based on experimental data, where we can investigate the consequences of injuries, alterations, resection surgery, etc. 
It seems that the application of these ideas to the brain are maturing, judging by the impact that scientific reports on the topic  receives, by the appearance of new books condensing the results of different laboratories, and by the increasing number of scientific meetings devoted to the subject of criticality in the brain. Although a ``theory of the brain'' is extremely far away, we believe that the transfer of methods of statistical physics to the sciences of the brain is moving us in that direction. 

\section{Critical fluctuations in the native state of proteins} 
As discussed so far, the current perspective views the brain as emerging from the interaction of an astronomical number of neurons, and able to adopt different phases. In the same sense the complex interaction of thousands of proteins shares information and are the basis of cell metabolism,  the central and fundamental requirement for life.  In this section we will discuss how some relevant aspects of this problem can be related to critical phenomena.

Proteins are unbranched chains of amino acids with different conformations, where the globular type is the most frequent in nature.  The  three-dimensional folded structure, known as native states, makes proteins capable of performing their biological functions. 
A globular protein carries out its functions by switching from one structural conformation  to another, even transiently, for instance when it recognizes and binds with other molecules. To achieve such performance, the structure of the native state of the protein must be susceptible enough to sense the signal and flexible to switch to another structure, but also be stable enough to warrant functional specificity and structural robustness.
\begin{figure}[!ht]
\begin{center}
\includegraphics[width=.95\linewidth]{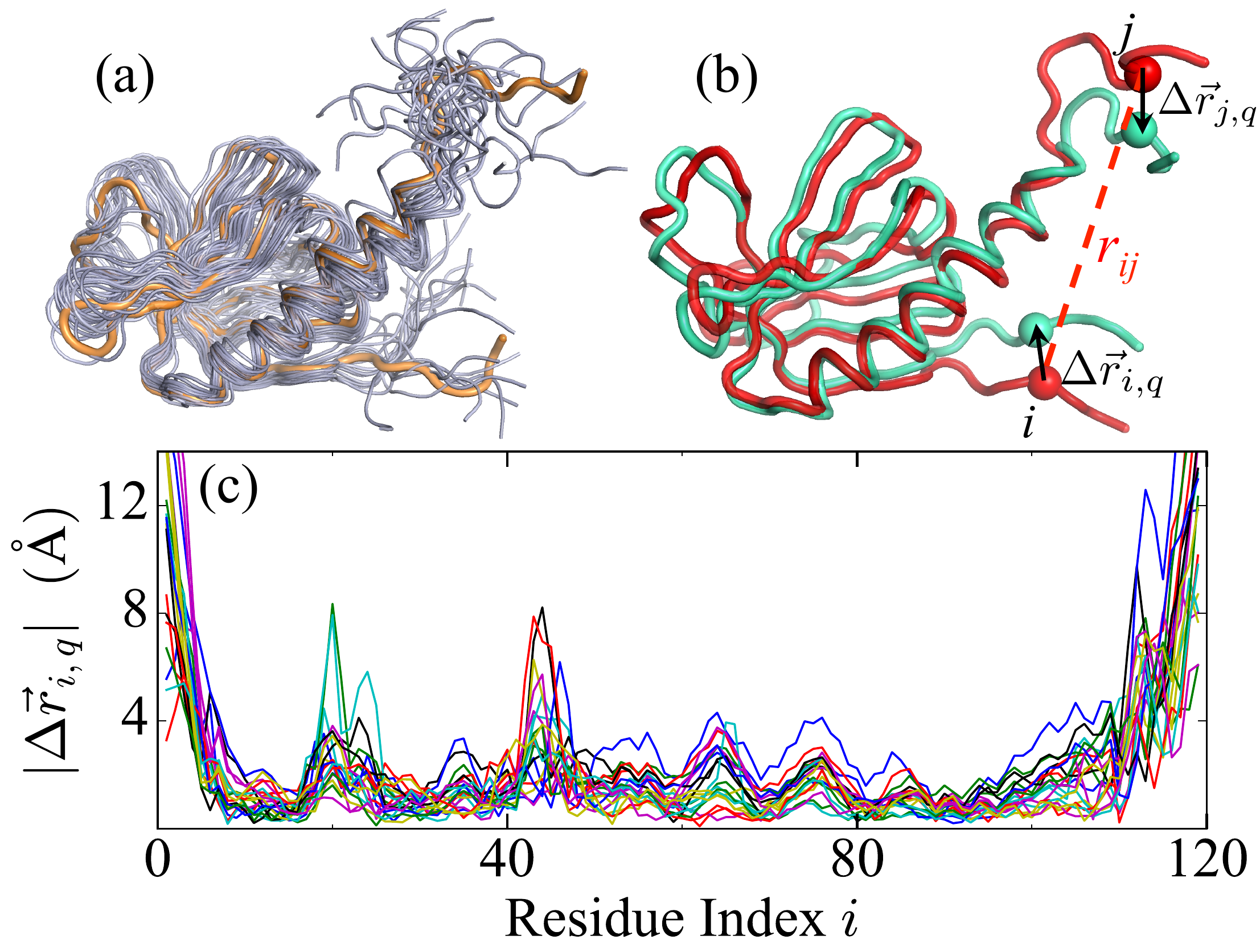}
\caption{(a) Example of a structural ensemble (protein PDB code: 1BAK) showing the  20 structures (grey) aligned to the reference one (orange).
(b) The average structure (red) and the $q$-th structure (cyan) for the protein in (a). The displacement $\Delta \vec{r}_{i,q}$ (or $\Delta \vec{r}_{j,q}$) of the $i$-th (or $j$-th) residue from its counterpart in the average structure is marked with black arrows. The distance $r_{ij}$ between residues $i$ and $j$ is marked with red dashes. (c) The magnitude of residues' position fluctuations $|\Delta\vec{r}_{i,q}| $ for the $q$-th structure ($q = 1, 2, \cdots$) in the ensemble. Adapted from \cite{tang}.}
\label{cartoon}
\end{center}
\end{figure}

From the arguments exposed in the previous section it  should be clear that the protein native state cannot be too rigid neither too flexible. The required  intermediate levels of flexibility for the proteins to function are  very similar to the properties seen in the vicinity of a critical point.  When proteins fold they undergo a phase transition from state that has high entropy and high free energy to a state os low entropy and low free energy. The question here is whether the  native state shares the generic properties of a critical point. \cite{bak, mora, reviewBiophys, chialvo, chate}

Critical fluctuations in protein equilibrium dynamics has been emphasized already by a number of results, including the power-law relation between solvent-accessible surface area and volume of proteins \cite{moret}, the fractal-like structure of configuration space \cite{smith2008} and oscillation spectrum \cite{reuveni2008}, the slowness of relaxation in protein molecules \cite{lu, smith2016}, the overlap between the low-frequency collective oscillation modes and large-scale conformational changes in allosteric transitions \cite{bahar1998, bahar2010, yang}, critical water fluctuation near hydrated proteins \cite{chalmers}, and so on. A few had already adventured to call the native state an example of self-organized criticality as in the work by Phillips \cite{phillips} or in the discussion on pairwise correlations between residues in protein families \cite{mora}. Yet, a direct characterization of the critical fluctuations near the native states of proteins based on experimental data was incomplete.  
\begin{figure}[htbp]
\begin{center}
\includegraphics[width=.97\linewidth]{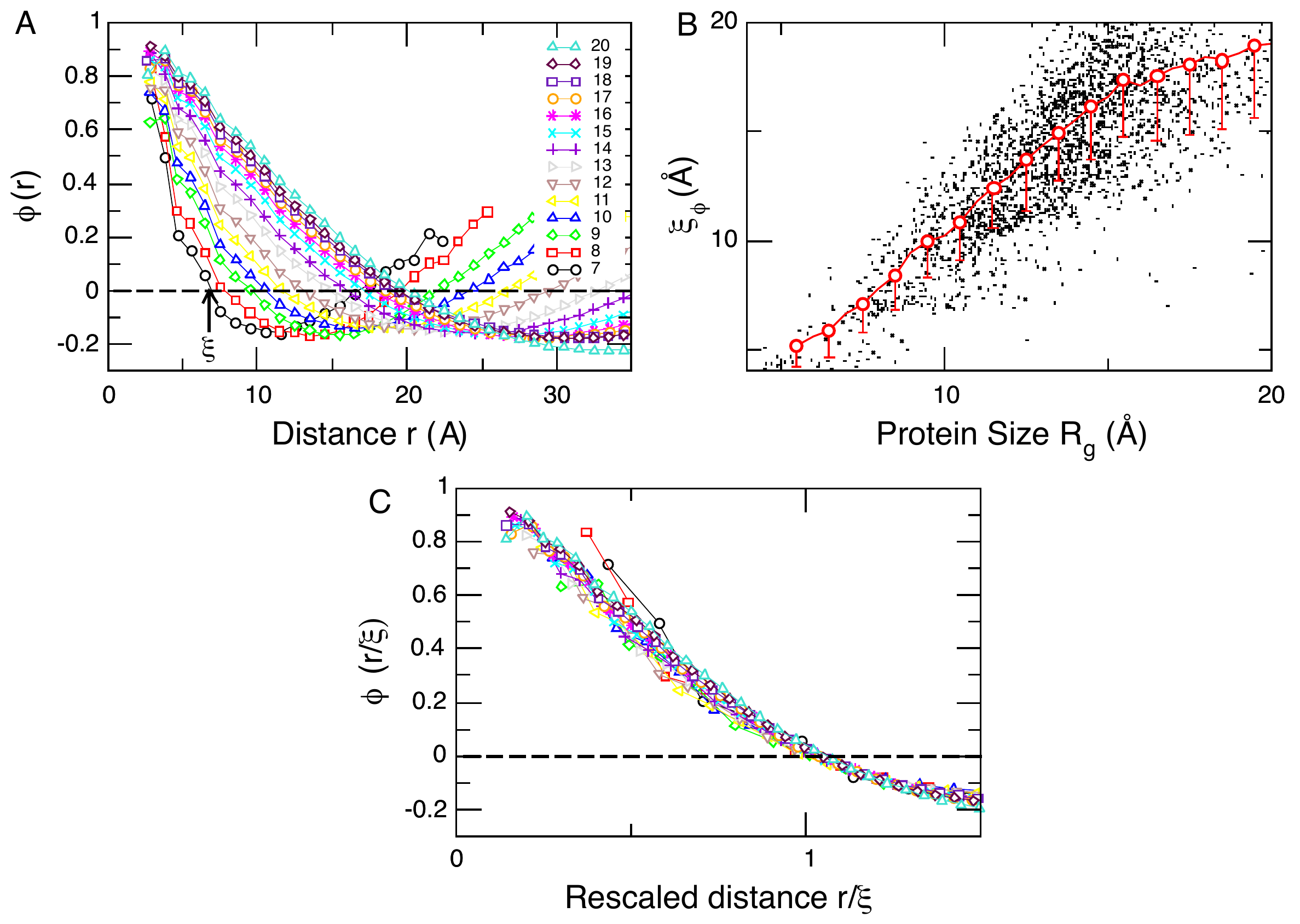}
\caption{\label{protein2}
(A) The distance-dependent correlation function $\phi(r)$  for proteins of different sizes (averaged over the $R_g$ values denoted in the legend). As an example the arrow indicates  $\xi$ for the smallest proteins. (B) Scattering plot of correlation length $\xi_\phi$ as a function of the average $R_g$, where the red symbols show the average $\xi= f(R_g)$ with the error bars denoting the standard deviation. 
(C) The scaling plot of $\phi(r/\xi)$. Figure redrawn from \cite{tang}.
}
\end{center}
\end{figure}
Recently we have studied the fluctuations around the native states of a large number of proteins based on their structural ensembles determined by solution nuclear magnetic resonance (NMR). 
For this analysis each structure of the ensemble is considered as one instantiation of the many conformations that the protein can adopt in the native basin.  
In order to test the conjecture of criticality we examined the distance-dependent correlations of position fluctuations of residues.  A large database of thousands of proteins structures was curated  and used to conduct finite size scaling analysis of the correlation functions (see details in \cite{tang}). The working hypothesis was that the correlations and susceptibility scales with the size of the proteins,  features that are similar to those in other physical systems near their critical points. This will imply that even weak local perturbations to any given residue can be felt by every other residue of the entire protein.

In Fig.\ref{cartoon} an example of the type of data for one protein is presented together with relevant notation: each ensemble includes $Q$ realizations of the same protein molecule, made up of $N$ amino acids with its coordinates denoted as $\vec{r_i}=[x_i,y_i,z_i]$. The fluctuations of each residue' coordinates ($\vec{\Delta r_i}$) across the ensemble are defined as, 
\begin{equation}
\vec{\Delta r_i}= \vec{r_i} - \frac{1}{Q} \sum_{q = 1}^Q \ \vec{r_{i,q}} 
\end{equation}
or alternatively by their magnitude by taking the norm of $\vec{\Delta r_i}$. 
Fig.\ref{cartoon}(C) shows the typical fluctuations exhibited by $\vec{\Delta r_i}$ in one ensemble of $Q=20$ structures.
The correlation properties of these fluctuations can be described, following similar calculations in Ref.\cite{cavagna2010, attanasi}, by the distance-dependent correlation function,
\begin{equation}
C(r)=\frac{\sum_{i\neq j}^N \ \vec{\Delta r_i} \cdot \vec{\Delta r_j}\ \delta(r-r_{ij})}{\sum_{i\neq j}^N \ \delta(r-r_{ij})} \
\end{equation}
where $\delta(r-r_{ij}) $ is a smoothed Dirac delta-function selecting pairs of residues at mutual distance $r$. 
Normalizing the covariance for residue pairs in the usual way, we have  $\phi_{ij}= \left( \Delta \vec{r}_i \cdot \Delta \vec{r}_j \right) / \sqrt{ \left( \Delta \vec{r}_i \cdot \Delta \vec{r}_i \right)\cdot \left( \Delta \vec{r}_j \cdot \Delta \vec{r}_j \right) }$, which (written in terms of mutual distances $r$) corresponds to the functions $\phi(r)$ plotted in Fig.\ref{protein2}A.
Similar calculations apply to the case of taking the norm of $\vec{\Delta r_i}$. 
\begin{figure}
\begin{center}
\includegraphics[width=.95\linewidth]{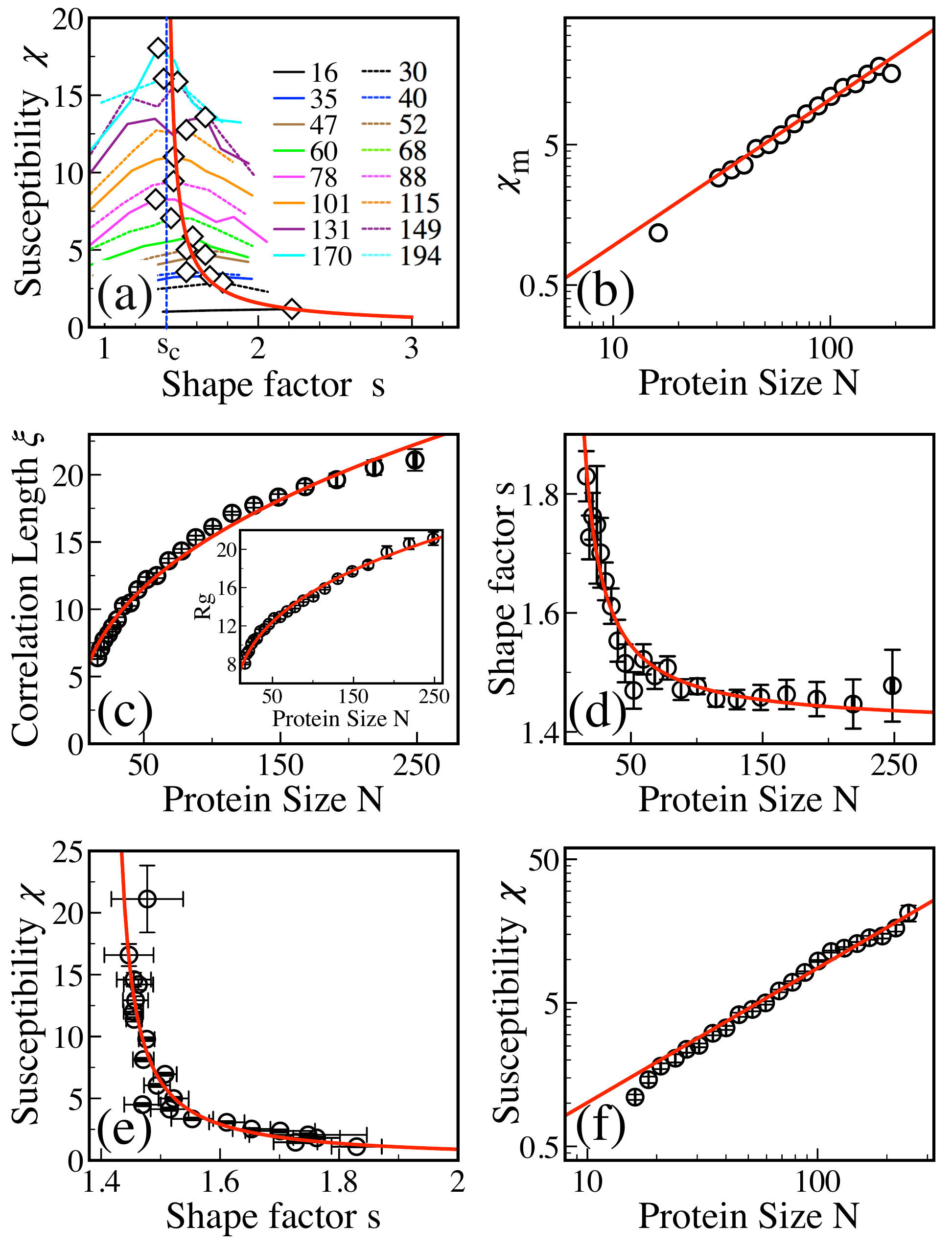}
\caption{Proteins finite size scaling. (a) Susceptibility $\chi$ for sets of proteins with different $N$ as a function of the control parameter $s$.  The peak heights $\chi_m$ and positions $s_m$ of susceptibility show a power-law relation $\chi_{m} \sim (s_{m}-s_c)^{-\gamma}$ (the red thick line). (b) The peak heights scaled with $N$ as $\chi_{m}\sim N^{-\alpha\gamma/\mu}$. 
(c) Correlation length $\xi$ as a function of protein size $N$. The inset shows $R_g$ as a function of $N$.
(d) Control parameter $s$ as a function of $N$.  (e) Susceptibility $\chi$ as a function of control parameter $s$. (f) Susceptibility $\chi$ as a function of the number of residues $N$. 
In (c)-(f), error bars represent the standard error of the data.}
\label{protein3}\end{center}
\end{figure}
Each curve in  Fig.\ref{protein2}A correspond to the distance-dependent correlation function averaged over proteins of a given giration ratio (indicated in the legend). The zero crossing of the curves is the correlation length $\xi$ which clearly increases with protein size (as seen in panel B for all proteins studied). The correlations functions of Fig.\ref{protein2}A are reploted in Panel C after rescaling the distance by $\xi$. This rescaling produce a good collapse of all curves, revealing the fact that there is a single scale able to  describe the correlation properties of all thousand of proteins. 
\begin{figure} [!hb]
\centering
\includegraphics[width=.95\linewidth]{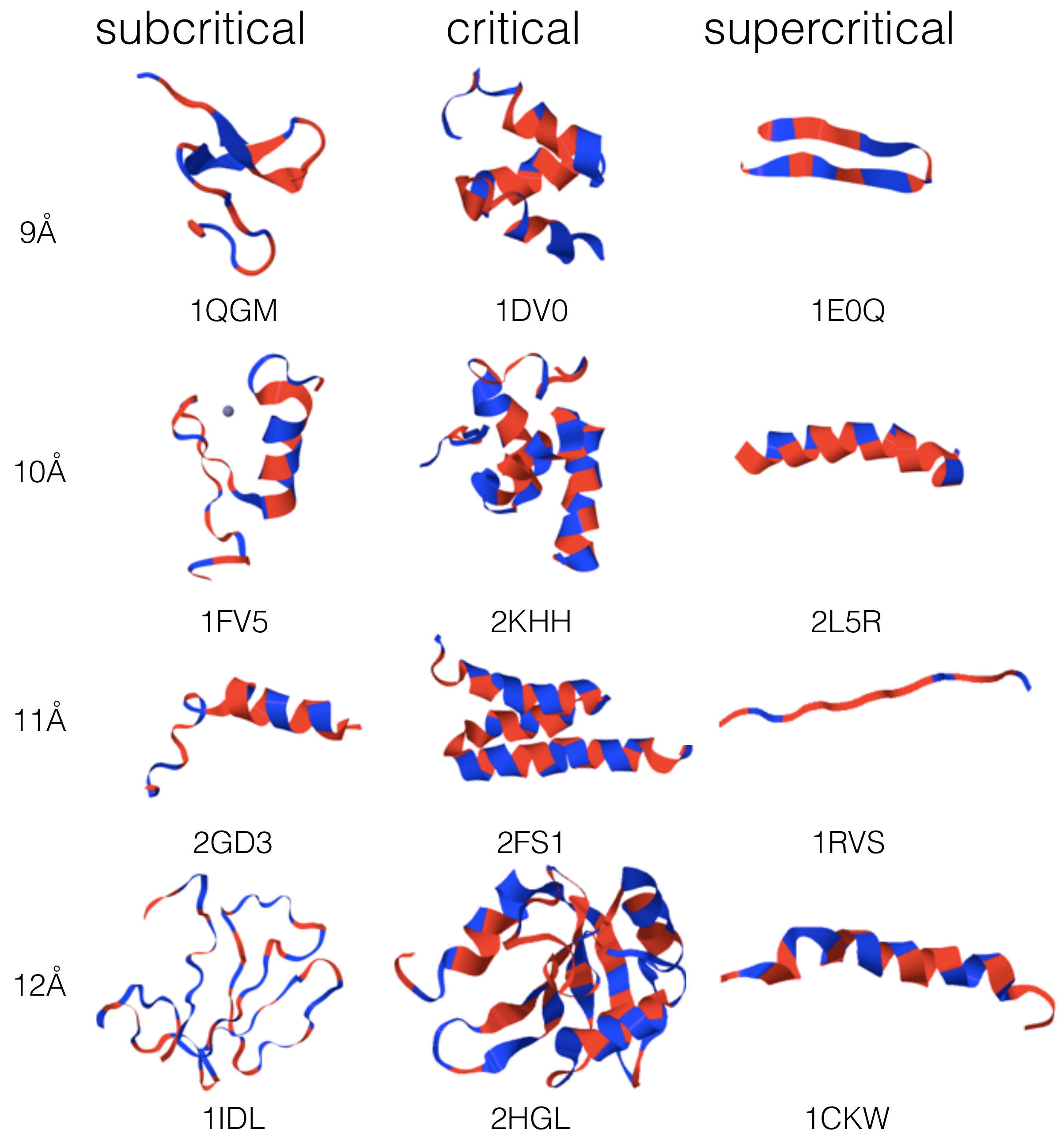}
\caption{\label{protein4}Typical examples of subcritical, critical and supercritical proteins for different sizes (gyration ratios indicated on the left).}
\end{figure}
If the increase of $\xi$ with protein size is due to critical phenomena it can be investigated by conducting a simple finite size scaling analysis as was done in Ref. \cite{attanasi}.
Thus, we tested such possibility for proteins of different sizes. To start, for each protein, the susceptibility was computed as the summed correlation between residue pairs, that is:
\begin{equation}
\chi = \frac{1}{N} \sum_{i\neq j}^N \phi_{ij} \cdot \theta(\xi_\phi - r_{ij}).
\end{equation}

As a pseudo-control parameter of the proteins we choose  a dimensionless shape factor $s$  defined as $s = Na^3/(L_aL_bL_c)$, where $a=3.8\AA$ is the size of a residue and $L_a$, $L_b$ and $L_c$ are the lengths of the principle axis of the protein ($L_a \leq L_b \leq L_c$).  For sphere-like protein molecules, the value of $s$ is relatively large (densely packed, and solid-like), while for elongated chains (loosely packed, and polymer-like), $L_c=Na$, and $L_a=L_b = a$, thus $s = 1.$ 
If the results correspond to critical behavior then the following relations are expected to hold:
 
\begin{equation}
 \xi \sim N^{\alpha}~~, ~~ 
 s-s_c \sim N^{-\alpha/\nu}~~, ~~
\chi \sim N^{\alpha\gamma/\nu} 
 \end{equation}

The results revealed scaling behaviors as shown by the fittings (red lines in panels of Fig.\ref{protein3}(a-e)) from which the exponents are determined. (1) Based on the relation $\xi \sim N^{\alpha}$, we get $\alpha = 0.40$ (Fig.\ref{protein3}(b)), which is similar to the result $\alpha=0.32$ based on $R_g \sim N^{\alpha}$ (inset of Fig.\ref{protein3}(b)). This indicates that proteins are tightly packed, and is consistent with the critical shape factor $s_c$ \cite{sc}. (2) For $s-s_c \sim N^{-\alpha/\nu}$, we have $1/\nu \approx 2.87$ (or $\nu \approx 0.35$) (Fig.\ref{protein3}(c)). (3) $\gamma$ can be determined based on the relations between $\chi$ and $s$ (or $N$). Fig.\ref{protein3}(d) shows the relation $\chi \sim (s-s_c)^{-\gamma}$ with $\gamma \approx 1.05$, and Fig.\ref{protein3}(e) depicts the relation $\chi \sim N^{\alpha\gamma/\nu}$ with $\gamma \approx 1.03$ which are comparable to the fitting result ($\gamma \approx 1.01$) in Fig.\ref{protein3}(a). 
Being conservative and taking integers, the approximated exponents were $\alpha = 1/3$, $\nu = 1/3$, $\gamma = 1$. 

These scaling relations are signatures of the critical character of the fluctuations of proteins and resemble scale-free correlations seen in other systems near critical points.
 
\begin{figure} [!hb]
\centering
\includegraphics[width=.9\linewidth]{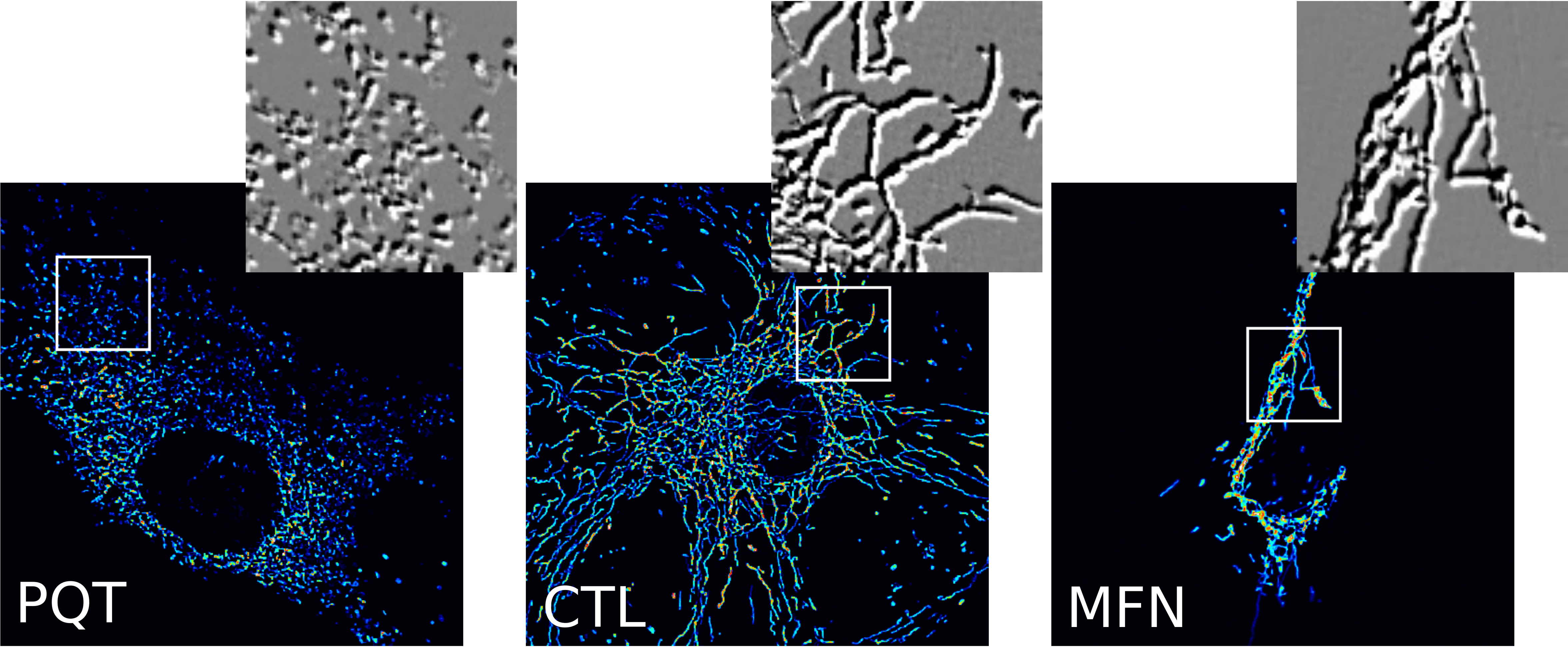}
\caption{\label{mito1}Typical examples of mitochondrial networks extracted under control (CTL) conditions, as well as with the two manipulations: over-expression of mitofusin1 (MFN) and treatment with paraquat (PRQ). The insets correspond to the skeletons extracted from the boxed regions (box side =$15 \mu$m). The color scale depicts the fluorescence intensity of mYFP. Adapted from \cite{nahuel1}}
\end{figure}

\section{Mitochondrial network complexity emerges from critical fission/fusion balance}

Now we turn our attention to describe a very complex cell organelle. Mitochondria arose around two billion years ago from the engulfment of an $\alpha$-proteobacterium by a precursor of the modern eukaryotic cell \cite{evolution}. Subsequent evolution shaped the relation between mitochondria and its host cell, leading to a highly specialization of both morphology and function of this organelle. Long known for its role in ATP production, mitochondria also participates in a myriad of processes such as apoptosis, calcium buffering and phospholipid synthesis, among others \cite{formfunct}. In addition, complex dynamic patterns occur in mitochondria, including oscillations and phase transitions \cite{oscillations, transition}. Thus, it is not surprising that a variety of functional alterations impact on mitochondrial morphology and viceversa. Although it is accepted that the structural status of the network is a predictor of the functional state of the organelle, only recently detailed quantitative studies of this relation are appearing, mostly due to the inherent difficulty in estimating  changes in its complex structure.

A typical mitochondria comprises a complex network of tubule-like structures, with fragments of all sizes (ranging from less than $1\mu m$ to $15 \mu m$ or more) \cite{morpho}.  The current theoretical understanding propose that mitochondrial network morphology is maintained by two opposing processes, fission and fusion, which depending on their relative predominance determine the overall network shape and morphological properties \cite{mfncite1}. 
\begin{figure} [!hb]
\centering
\includegraphics[width=.9\linewidth]{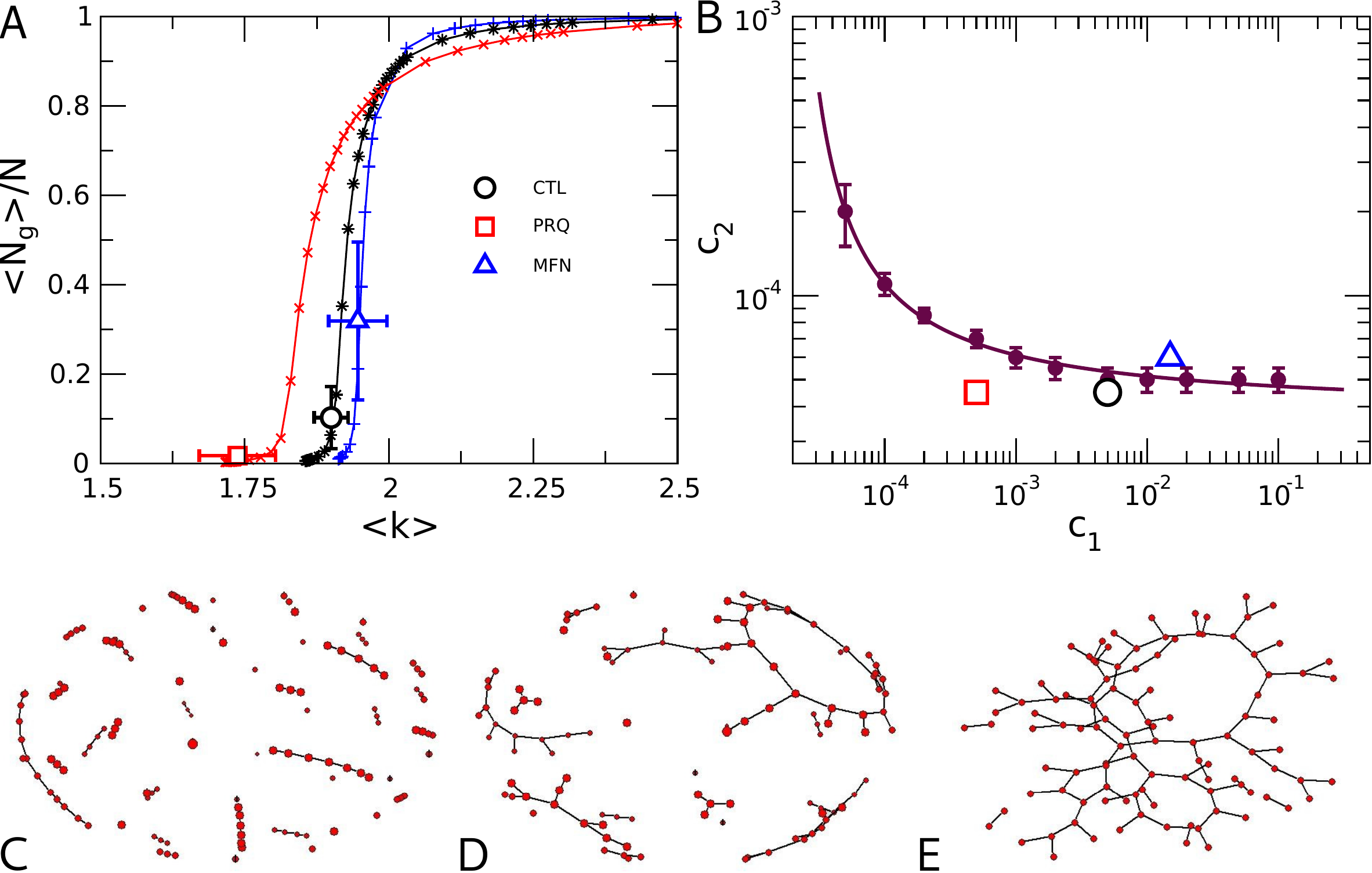} 
\caption{\label{mito2}Comparison of the present experimental results with those of Sukhorukov' {\it et al.} model. (A) Model parameters extracted iteratively from the experimental data. Open symbols and error bars correspond to means and standard deviations of all cells for the CTL, PRQ and MFN groups (for a binarization threshold = 0.15). Each point of the extracted curves corresponds to a unique pair of values $(c_1,c_2)$.
(B) Phase diagram of the model in the $(c_1,c_2)$ space. Filled symbols and the continuous line correspond to the location of the phase transition. The three open symbols (labeled PRQ, CTL, MFN) correspond to the parameter values extracted from the experimental data as shown in (A). Panels C, D and E show a graphical representation of the typical networks simulated using the three derived $(c_1,c_2)$ values. Adapted from \cite{nahuel1}}
\end{figure}

On recent work Zamponi et al. \cite{nahuel1} proposed that the physiological structure of the mitochondria should be maintained on a critical balance in between these extremes. Fig.\ref{mito1} shows typical examples of mitochondria network for the control (CTL) case as  well as for two manipulations that either fragment (by action of paraquat, PRQ) or fuse the network by over-expression of mitofusin1 (MFN), modification that alter these opposing processes, namely extensive fission or fusion. In order to study  these changes quantitatively we defined a graph after extracting the skeleton of each network (insets in Fig.\ref{mito1}) and proceeded to study its topological properties. 
We found that mitochondrial morphology in control cells exhibited distinctive scale-free features, which were disrupted by both experimental manipulations in opposite directions. These observations were consistent with predictions of the Sukhorukov \textit{et al.} model \cite{Sukhorukov} of network growth where healthy mitochondrial networks are in a critical regime intermediate between the fragmented and fussioned networks.

The results on Fig.\ref{mito2} shows a comparison (Panel A) between the empirical estimation of network features, the normalized size of the giant cluster $N_g/N$ and the average degree $<k>$ of the network with the Sukhorukov \textit{et al.} model  predictions for the three conditions explored  experimentally.  In Panel B of the same figure the averages for the three conditions studied are plotted together with the critical line predicted in the Sukhorukov \textit{et al.} model  where it can be seen that the control condition lies near the critical line and in between the fragmented and fused cases. Finally, the panels C-E show simulated networks using the average parameters from the experiments.

\section{Summing up}

In these notes, we have discussed first how complexity in nature is often discovered in between the extremes of very ordered and very disordered regimes. Then we called attention to the fact that the well known scenario of a second order phase transition naturally includes such intermediate region of high complexity. That leads us to present results from three biological examples in which complex dynamics and structures are explained by the presence of critical phenomena. 

In closing we shall comment that each of the three examples presented here, have been already replicated in very different settings. Recent experiments by Scott \& colleagues \cite{leech} have faithfully replicated the phase transition we uncovered in fMRI experiments, in their case in optogenetics recordings from mice experiments. Concerning proteins dynamics, Fabio et al. \cite{posterAFA} recently conducted extensive molecular dynamics simulations of a subset of the database studied in Ref \cite{tang} and was able to demonstrate similar scaling properties.  Finally the approach presented here to study  the mitochondria network status has been used successfully to restore mitochondrial structure and function in Down syndrome cells \cite{nahuel2}.   
It is remarkable how universality allows to use the exact same framework to study complex phenomena of very different nature and scales, from a culture of few thousand neurons to the entire brain, from a small protein molecule to a network spanning the entire cell.
\\
  
{\it Acknowledgements:} DRC thanks the hospitality of the Jagellonian University where part of this work was conducted while receiving partial 
support from the Jagellonian University-UNSAM Cooperation Agreement and from the MAESTRO DEC-2011/02/A/ST1/00119 grant of the National Center of Science (Poland). DRC is also funded by the grant 1U19NS107464-01 from the NIH (USA).


\begin{thebibliography} {50} 
\bibitem{glass} Glass L, Mackey MC. (1988) From clocks to chaos: The rhythms of life, Princeton University Press. 
 
\bibitem{chialvo} Chialvo DR. (2010) Emergent complex neural dynamics. {\it Nature Physics} {\bf 6}, 744. 


\bibitem{tang}Tang QY, Zhang YY, Wang J, Wang W, Chialvo DR. (2017) Critical fluctuations in the native state of proteins. {\it Phys. Rev. Letters}  {\bf 118}(8), 088102.

\bibitem{nahuel1} Zamponi N, Zamponi E, Cannas SA, Billoni OV, Helguera PR, Chialvo DR. (2018) Mitochondrial network complexity emerges from fission/fusion dynamics. {\it Scientific Reports} {\bf  8}, 363.  

 
\bibitem{hans} Frauenfelder, H. (1987) Function and dynamics of myoglobin. {\it Annals of the New York Academy of Sciences} {\bf 504}, 151--167. 
 
\bibitem{bak} Bak P. (1996) How nature works: The science of self-organized criticality. New York: Springer Science. 
  
\bibitem{beggs} Beggs JM \& Plenz D. (2003) Neuronal avalanches in neocortical circuits. {\it Journal of Neuroscience}  {\bf 23}, 11167. 
  
 \bibitem{mentecritica} Marro J \& Chialvo DR. (2017) La mente es critica. Descubriendo la admirable complejidad del cerebro. Granada: Editorial University of Granada. 
 
\bibitem{PP1} Tagliazucchi E, Balenzuela P, Fraiman D, Montoya P, Chialvo DR. (2011). Spontaneous BOLD event triggered averages for estimating functional connectivity at resting state. {\it Neurosci. Lett.}  {\bf 488}(2), 158--163.

\bibitem{PP2} Tagliazucchi E, Siniatchkin M,  Laufs H, Chialvo DR. (2016). The voxel-wise functional connectome can be efficiently derived from co-activations in a sparse spatio-temporal point-process. {\it Front. Neurosci.} /doi.org/10.3389/fnins.2016.00381  

\bibitem{PP3} Tagliazucchi E, Balenzuela P,  Fraiman D, Chialvo DR. (2012) Criticality in large-scale brain fMRI dynamics unveiled by a novel point process analysis. {\it Frontiers in Physiology}  {\bf 3}, 15. 

\bibitem{fraiman} Fraiman D \& Chialvo DR. (2012) What kind of noise is brain noise: anomalous scaling behavior of the resting brain activity fluctuations. {\it Frontiers in physiology} {\bf 3}, 307.

\bibitem {tononi} Tononi G, Boly M, Massimini M, Koch C. (2016) Integrated information theory: from consciousness to its physical substrate. {\it Nature Reviews Neuroscience}  {\bf 17}, 450--461. 

\bibitem {tononi2} Tononi G. (2004) An information integration theory of consciousness. {\it BMC Neuroscience}  {\bf 5} (1):42.

 \bibitem{edge} Chialvo DR. (2007) The brain at the edge. In ``Cooperative Behavior in Neural Systems: Ninth Granada Lectures'', J. Marro, P. L. Garrido, \& J. J. Torres (Eds). 

\bibitem{Tegmark} Tegmark M. (2015) Consciousness as a state of matter.  {\it Chaos, Solitons \& Fractals} {\bf 76}, 238--270.

\bibitem{tegmark1} Hut P, Alford M, Tegmark M. (2006) On math, matter and mind. {\it Found. Phys.} {\bf 36}, 765-794.



\bibitem{propofol}Tagliazucchi E, Chialvo DR, Siniatchkin M, Amico E, Brichant  JF. (2016) Large-scale signatures of unconsciousness are consistent with a departure from critical dynamics. {\it Journal of the Royal Society Interface } {\bf 13} (114), 20151027

\bibitem{hongos} Tagliazucchi E, Carhart-Harris RL, Leech R, Nutt D, Chialvo DR. (2014) Enhanced repertoire of brain dynamical states during the psychedelic experience. {\it Human brain mapping } {\bf 35} (11), 5442--5456.

 
\bibitem{robin}Carhart-Harris RL, Leech R, Hellyer PJ, Shanahan M, Feilding A. (2014). The entropic brain: a theory of conscious states informed by neuroimaging research with psychedelic drugs.  {\it Frontiers in Human Neuroscience} {\bf 8}, 20.

\bibitem{haimovici} Haimovici A, Tagliazucchi E, Balenzuela B and Chialvo DR. (2012) Brain organization into resting state networks emerges at criticality on a model of the human connectome. {\it Phys. Rev. Letters } {\bf 110}, 178101.

\bibitem{connectome} Human Connectome Project website: http://www.humanconnectomeproject.org 

\bibitem{rocha}Rocha RP,  Kocillari L, Suweis S, Corbetta M,  Maritan A. (2018). Homeostatic plasticity and emergence of functional networks in a whole-brain model at criticality {\it Scientific Reports} {\bf 8} (1), 15682.

\bibitem{mora}Mora T  \& Bialek W. (2011) Are biological systems poised at criticality? {\it J. Stat. Phys.}  {\bf 144}, 268.

\bibitem{reviewBiophys} Honerkamp-Smith AR, Veatch SL, Keller SL. (2009) An introduction to critical points for biophysicists; observations of compositional heterogeneity in lipid membranes. {\it Biochim. Biophys. Acta} {\bf 1788}, 53.

 
\bibitem{chate} Chat\'e  H  \& Mu\~noz M. (2014) Insect swarms go critical. {\it Physics} {\bf 7}, 120.

\bibitem{moret}Moret MA \& Zebende GF. (2007) Amino acid hydrophobicity and accessible surface area. {\it Phys. Rev. E} {\bf 75}, 011920.

\bibitem{smith2008} Neusius T, Daidone I, Sokolov IM, Smith JC. (2008) Subdiffusion in peptides originates from the fractal-like structure of configuration space. {\it Phys. Rev. Lett.} {\bf 100}, 188103.

\bibitem{reuveni2008} Reuveni S, Granek R, Klafter J. (2008) Proteins: coexistence of stability and flexibility. {\it Phys. Rev. Lett.} {\bf 100}, 208101.

\bibitem{lu}Lu HP, Xun L, Xie XS. (1998) Single-molecule enzymatic dynamics. {\it Science} {\bf 282}, 1877--1882.

\bibitem{smith2016} Hu X, Hong L, Smith MD, Neusius T, Cheng X, and Smith JC. (2016). The dynamics of single protein molecules is non-equilibrium and self-similar over thirteen decades in time.{\it  Nat. Phys.} {\bf 12}, 171.

\bibitem{bahar1998} Bahar I, Atilgan AR, Demirel MC, Erman B. (1998) Vibrational dynamics of folded proteins: significance of slow and fast motions in relation to function and stability. {\it Phys. Rev. Lett.} {\bf 80}, 2733.

\bibitem{bahar2010} Bahar I,  Lezon TR, Yang LW, Eyal E. (2010) Global dynamics of proteins: bridging between structure and function. {\it  Annu. Rev. Biophys.}  {\bf 39}, 23--42.

\bibitem{yang}Yang L, Song G, Jernigan RL. (2007) How well can we understand large-scale protein motions using normal modes of wlastic network models? {\it Biophys. J. } {\bf 93}, 920--929.

\bibitem{chalmers} Patel AJ, Varilly P, Jamadagni SN, Hagan MF, Chandler D, Garde S. (2012) Sitting at the Edge: How biomolecules use hydrophobicity to tune their interactions and function. {\it J. Phys. Chem. B } {\bf 116}, 2498--2503.

\bibitem{phillips} Phillips JC. (2009) Scaling and self-organized criticality in proteins I.  Proc. Natl. Acad. Sci. U.S.A. {\bf 106}, 3107-3112; Phillips JC (2009) Scaling and self-organized criticality in proteins II {\it Proc. Natl. Acad. Sci. U.S.A.} {\bf 106}, 3113-3118.

\bibitem{best2006} Best RB, Lindorff-Larsen K,  DePristo MA, and Vendruscolo M. (2006) Relation between native ensembles and experimental structures of proteins {\it Proc. Natl. Acad. Sci. U.S.A.} {\bf 103}, 10901--10906.

\bibitem{palmer} Palmer  AG 3rd. (2004) NMR characterization of the dynamics of biomacromolecules. {\it Chem. Rev.} {\bf 104}, 3623-3640.

\bibitem{lezon2010} Lezon TR \& Bahar I. (2010) Using entropy maximization to understand the determinants of structural dynamics beyond native contact topology. {\it PLoS Comput. Biol.} {\bf 6}, e1000816.

\bibitem{Lammert} Lammert H, Noel JK, Haglund E, Schug A, Onuchic JN. (2015) Constructing a folding model for protein S6 guided by native  fluctuations deduced from NMR structures {\it J. Chem. Phys.} {\bf 143,} 243141.


\bibitem{cavagna2010} Cavagna A, Cimarelli A, Giardina I, Parisi G, Santagati R, Stefanini F, Viale M. (2010) Scale-free correlations in starling flocks {\it Proc. Natl. Acad. Sci. U.S.A.} {\bf 107}, 11865-11870.

\bibitem{attanasi}Attanasi A, Cavagna A, Del Castello L, Giardina I, Melillo S, Parisi L, Pohl O, Rossaro B, Shen E, Silvestri E, Viale M. (2014) Finite-Size scaling as a way to probe near-criticality in natural swarms. {\it Phys. Rev. Lett.} {\bf 113}, 238102.

\bibitem{sc} Considering the scaling relation of the dense packing: $R_g = R_0 N^{1/3},$ (Fig.3(b) inset), where $R_0 \approx 3.4$. So, the critical shape factor $s_c \approx a^3/R_0^3 = (3.8/3.4)^3 \approx 1.396$ .

\bibitem{evolution} Gray MW, Burger G, Lang BF. (1999) Mitochondrial evolution. {\it Science} {\bf 283}, 1476--1481.

\bibitem{formfunct} Friedman JR, Nunnari J. (2014) Mitochondrial form and function. {\it Nature} {\bf 505}, 335--343.

\bibitem{oscillations} Kurz FT, Derungs T, Aon MA, O'Rourke B, Armoundas AA.  (2015) Mitochondrial networks in cardiac myocytes reveal dynamic coupling behavior. {\it Biophysical Journal} {\bf108}, 1922--1933.

\bibitem{transition}Aon MA, Cortassa S, O'Rourke B.  (2004) Percolation and criticality in a mitochondrial network. {\it Proc Natl Acad Sci USA} {\bf101}, 4447--4452.

\bibitem{morpho} Collins, TJ, Berridge MJ, Lipp P, Bootman MD.  (2002) Mitochondria are morphologically and functionally heterogeneous within cells. {\it EMBO J} {\bf 21} 1616--1627.

\bibitem{mfncite1} Chen H \& Chan DC. (2004) Mitochondrial dynamics in mammals. {\it Curr Top Dev Biol} {\bf 59},119--144.

\bibitem{Sukhorukov} Sukhorukov VM, Dikov D, Reichert AS, Meyer-Hermann M.  (2012) Emergence of the mitochondrial reticulum from fission and fusion dynamics. {\it PLoS Comput Biol} {\bf 8} e1002745.


\bibitem{leech} Scott G, Fagerholm ED, Mutoh H, Leech R, Sharp DJ, Shew WL, Knopfel T. (2014) Voltage imaging of waking mouse cortex reveals emergence of critical neuronal dynamics. {\it The Journal of Neuroscience} {\bf 34}, 16611--16620. 

\bibitem{nahuel2}Zamponi E, Zamponi N, Coskun P, Quassollo G, Lorenzo A, Cannas SA, Pigino G,  Chialvo DR,  Gardiner K,  Busciglio J, Helguera P. (2018) Nrf2 stabilization prevents critical oxidative damage in Down syndrome cells. {\it Aging cell}, e12812. 

\bibitem{posterAFA} Fabio L, Asciutto EK, General IJ, Chialvo DR. (2018) Characterisation of protein native state fluctuations. Asoc. Fisica Argentina Anual Meeting. Buenos Aires. Argentina
 

\end{thebibliography}
\end{document}